\documentclass[10pt,twocolumn,twoside,letterpaper]{IEEEtran}

\usepackage{geometry}
\usepackage{subfigure}
\usepackage{mathtools}
\usepackage{paralist}       
\usepackage{textgreek}
\usepackage{graphicx}

\usepackage{siunitx}

\usepackage{textcomp}
\usepackage{gensymb}

\makeatletter
\def\ps@IEEEtitlepagestyle{
  \def\@oddfoot{\mycopyrightnotice}
  \def\@evenfoot{}
}
\def\mycopyrightnotice{
  {\footnotesize
  \begin{minipage}{\textwidth}
  \centering
  978-1-6654-2113-3/21/\$31.00 \copyright2021 IEEE
  \end{minipage}
  }
}

\usepackage{url}

\begin{document}

\title{\LARGE \bf
Fast Neutron Spectroscopy with a High-pressure Nitrogen-filled Large Volume Spherical Proportional Counter
}

\author{I. Giomataris$^{1}$, I. Katsioulas$^{2}$, P. Knights$^{2}$, I. Manthos$^{2,*}$, J.~Matthews$^{2}$, T. Neep$^{2}$, K. Nikolopoulos$^{2}$, T.Papaevangelou$^{1}$, B.Phoenix$^{2}$, R.Ward$^{2}$ 
\thanks{$^{1}$ 
IRFU, CEA, Universite Paris-Saclay, Gif-sur-Yvette, F-91191, France}
\thanks{$^{2}$School of Physics and Astronomy, University of Birmingham, B15 2TT,
United Kingdom} 
\thanks{*\tt\small i.manthos@bham.ac.uk}

\thanks{

Manuscript received December 2, 2021. This project has received funding from the European Union's Horizon 2020 research and innovation programme under the Marie Sk\l{}odowska-Curie grant agreements and no 845168 (neutronSphere) and no 841261 (DarkSphere). }}



\maketitle
\pagenumbering{gobble}

\begin{abstract}
We present a fast neutron spectroscopy system based on a nitrogen-filled, large volume gaseous detector, the Spherical Proportional Counter. The system has been successfully operated up to gas pressure of 1.5~bar. Neutron energy is estimated through measurement of the ${}^{14}$N(n,\textalpha)${}^{11}$B and  ${}^{14}$N(n,p)${}^{14}$C reaction products. These reactions have comparable cross sections and $Q$-values with the ${}^{3}$He(n,p)${}^{3}$H reaction making nitrogen a good alternative to ${}^{3}$He use for fast neutron detection. Two detectors were built at the University of Birmingham and are currently used for the measurement of fast and thermal neutrons in the University of Birmingham and the Boulby underground laboratory, respectively.  
\end{abstract}

\section{Introduction}

Since the discovery of the neutron, there have been continuous efforts for the development of a simple, affordable, and efficient fast neutron spectroscopy system; an invaluable tool for many scientific and industrial applications. However, such a system remains elusive and fast neutron measurements remain cumbersome, based on complex methods performed usually by specialised personnel~\cite{Brooks:2002oyj}.

The use of $^{3}$He based large volume detectors provide a possible solution. $^{3}$He is a non-flammable and non-toxic gas which also shows low \textgamma-ray efficiency and long lifetime, but its popularity along with the difficulty of production resulted to very high prices. Existing alternatives have several disadvantages~\cite{KOUZES20101035, Simpson2011} such as toxicity, poor efficiency, complicated response, and insufficient $\gamma$/n separation. The difficulty in fast neutron measurements is reflected in the scarcity of measured neutron spectra in many laboratories and industrial sites.

We propose the N$_{2}$SPHERE system~\cite{Katsioulas:2019iea}, aspiring to become a successful alternative to existing technologies. It is based on a nitrogen-filled large-volume proportional counter. The neutron energy is estimated by measuring the products of the $^{14}$N(n, \textalpha)$^{11}$B and $^{14}$N(n, p)$^{14}$C reactions. These reactions have comparable cross sections for fast neutrons to the $^{3}$He(n,p)$^{3}$He and ${}^{10}$B(n,p)$^{7}$L reactions. 
The use of a light element such as nitrogen keeps \textgamma-ray efficiency low, enhancing the signal to background ratio in mixed radiation environments.
\begin{figure}
\centering
\includegraphics[width=0.6\linewidth]{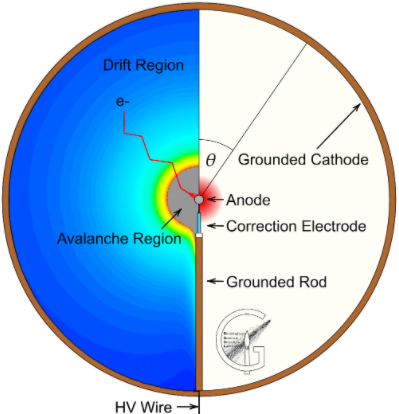}
\caption{The spherical proportional counter design and principle of operation.}
\label{fig:spc}
\end{figure}
\vspace*{-0.2cm}
\IEEEpubidadjcol
\section{The Detector of the N$_{2}$SPHERE System}
Operation of nitrogen-filled gaseous detectors at atmospheric pressure or above has been challenging due to low Townsend coefficient, requiring very high operation voltage. In addition, containing protons and $\alpha$-particles with $\mathcal{O}$(MeV) kinetic energy within the detector sensitive volume is challenging, requiring large volumes and high-pressure operation to minimise the wall-effect. 
As a result, ${}^{3}$He-filled cylindrical proportional counters are mainly used as flux monitors, moderating fast neutron energy, and rarely for spectroscopy. The use of the versatile Spherical Proportional Counter (SPC)~\cite{Giomataris_2008} aims to address these issues. The SPC comprises of a grounded spherical shell (cathode) and a small sphere (anode) placed at its centre, supported by a grounded metallic rod, to which high voltage is applied and from which the signal is read-out (Fig.~\ref{fig:spc}). The $1/{r}^{2}$ dependence, where $r$ is the radius from the centre, of the electric field results in large charge multiplication factors with anode diameters of $\mathcal{O}$(mm), and thus in a more robust and reliable detector construction compared to cylindrical counters. Details on the SPC and its advantages compared to other detector geometries can be found in Ref.~\cite{Katsioulas:2018squ}. 

Early explorations of this idea are discussed in Ref.~\cite{Bougamont:2015jzx}. This initial effort provided the proof of principle but suffered from issues such as wall effect, attachment and low charge collection efficiency, due to the early stages of the SPC development. In this work, we use the latest SPC instrumentation developments such as resistive multi-anode sensors~\cite{Giomataris:2020rna} for high-gain, high-charge collection efficiency and gas purifiers to minimize gas contaminants to negligible levels. More compact detectors (30-cm in diameter) are used, operating above atmospheric pressure, reducing the wall effect and increasing the sensitivity. 


%
%
\section{Neutron Measurements at the University of Birmingham}

\begin{figure}
\centering
\includegraphics[width=0.6\linewidth]{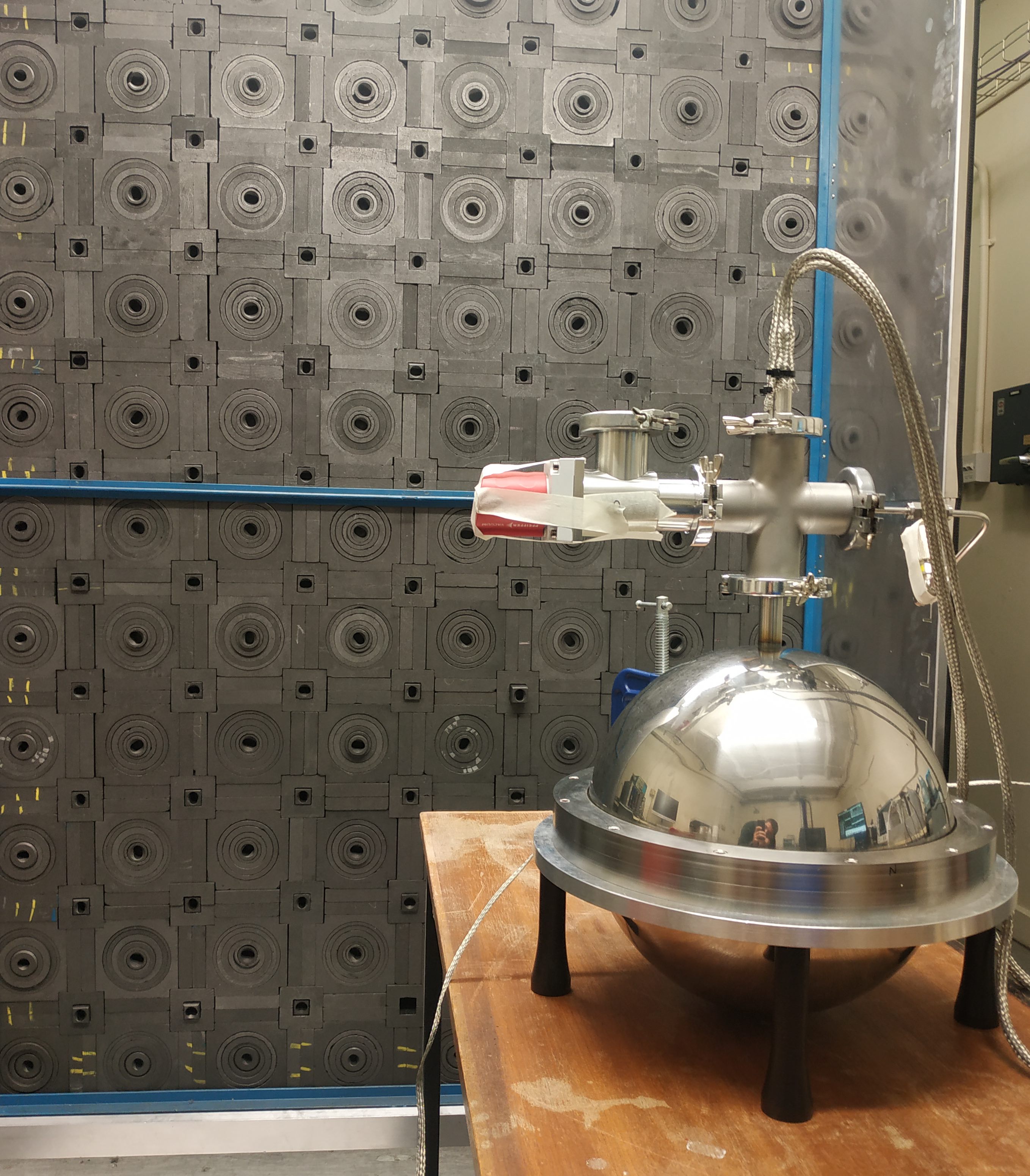}
\caption{Thermalised neutrons from $^{241}$Am$^9$Be, detected with a nitrogen-filled SPC operated at 1.5\,bar.}
\label{fig:graphiteStack_a}
\end{figure}

\begin{figure}[h] 
\centering
\includegraphics[width=0.85\linewidth]{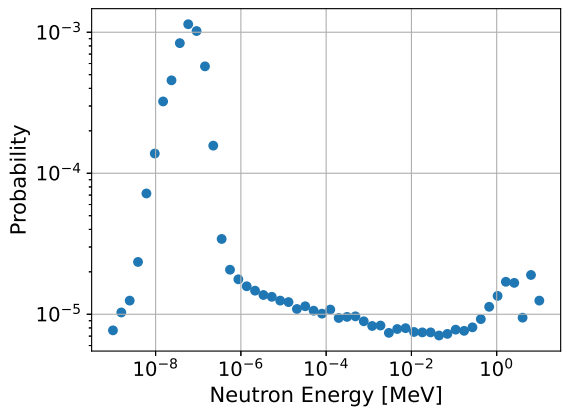}
\caption{The probability for a thermalised neutron to reach the detector volume.}
\label{fig:graphiteStack_b}
\end{figure}

An $^{241}$Am-$^{9}$Be neutron source, placed inside the graphite stack shown in Fig.~\ref{fig:graphiteStack_a}, is used to obtain thermal neutron measurements. Neutrons from the source thermalise through scattering in the graphite, and their energy spectrum estimated using dedicated simulations is shown in Fig.~\ref{fig:graphiteStack_b}. The probability of neutron from the source to reach the detector volume is estimated to be  $\approx5\times10^{-3}$.

\begin{figure}[h] 
\centering
\subfigure[\label{fig:3_6kV_a}]{\includegraphics[width=0.85\linewidth]{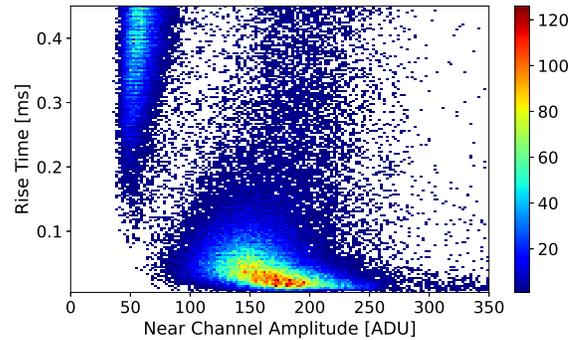}}
\subfigure[\label{fig:3_6kV_b}]{\includegraphics[width=0.85\linewidth]{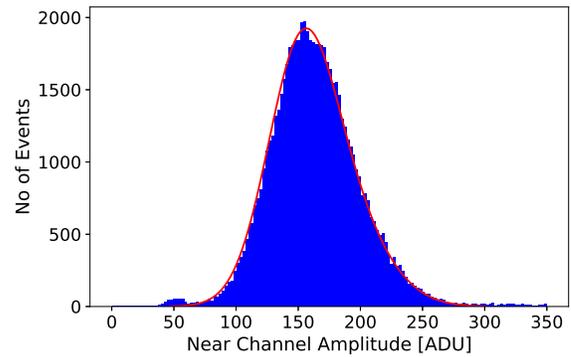}}
 \caption{Data at 1\,bar N$_{2}$ and 3.6\,kV anode voltage: \subref{fig:3_6kV_a} The rise time vs peak Amplitude after applying pulse shape quality criteria. \subref{fig:3_6kV_b} Thermal neutrons peak amplitude distribution. The peak corresponds to the 625\,keV recoil energy with 20.7\% energy resolution.\label{fig:3_6kV}}
\end{figure}

We used a 30-cm in diameter nitrogen-filled SPC to perform thermal neutron measurements in a variety of anode voltages, at pressures of 1\,bar and 1.5\,bar. The detector was equipped with an 11-anode sensor with 1\,mm anode diameter which was readout in two channels, the 5 anodes closest to the supporting rod (``near'' side), and the rest (``far'' side). Fig. \ref{fig:3_6kV_a} presents the pulse rise time versus its amplitude, while Fig. \ref{fig:3_6kV_b} the pulse amplitude distribution after applying pulse shape quality criteria (i.e. the FWHM), for 1\,bar pressure at 3.6\,kV anode voltage. The thermal neutrons peak is in correspondence with the alpha particles peak emitted by $^{222}$Rn decay, that is inserted to the detector volume by the gas purifier. A neutron efficiency of approximately $3.7\times10^{-4}$ is estimated. 

\begin{figure}[h] 
\subfigure[\label{fig:1_5bar_a}]{\includegraphics[width=.48\textwidth]{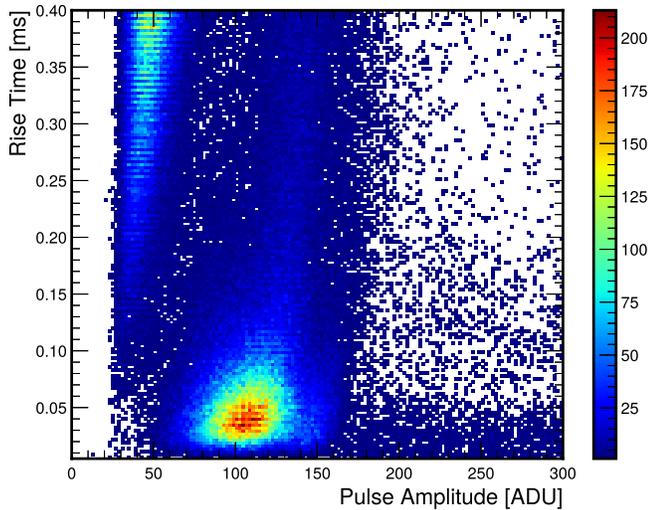}}
\subfigure[\label{fig:1_5bar_b}]{\includegraphics[width=.85\linewidth]{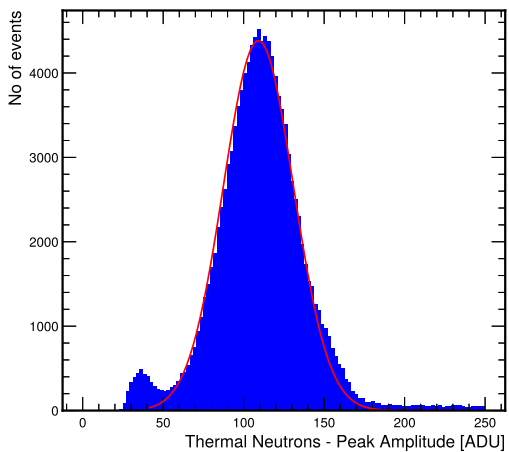}}
\caption{Data at 1.5\,bar N$_{2}$ and 4.5\,kV anode voltage: \subref{fig:1_5bar_a} The rise time vs peak Amplitude after applying pulse shape quality criteria. \subref{fig:1_5bar_b}  Thermal neutrons peak amplitude distribution. The peak at 109.16$\pm$0.16\,ADU corresponds to the 625\,kV recoil energy with a 19.93\% energy resolution.\label{fig:1_5bar}}
\end{figure}

The corresponding results for the operation of the detector at a pressure of 1.5\,bar with 4.5~\,kV anode voltage is shown in Fig. \ref{fig:1_5bar}. The thermal neutrons peak found in good agreement with the $\alpha$-particle peak produced by the $^{222}$Rn decay.

\section{Simulation Study}
\begin{figure}[!h]
\centering
\includegraphics[width=60mm]{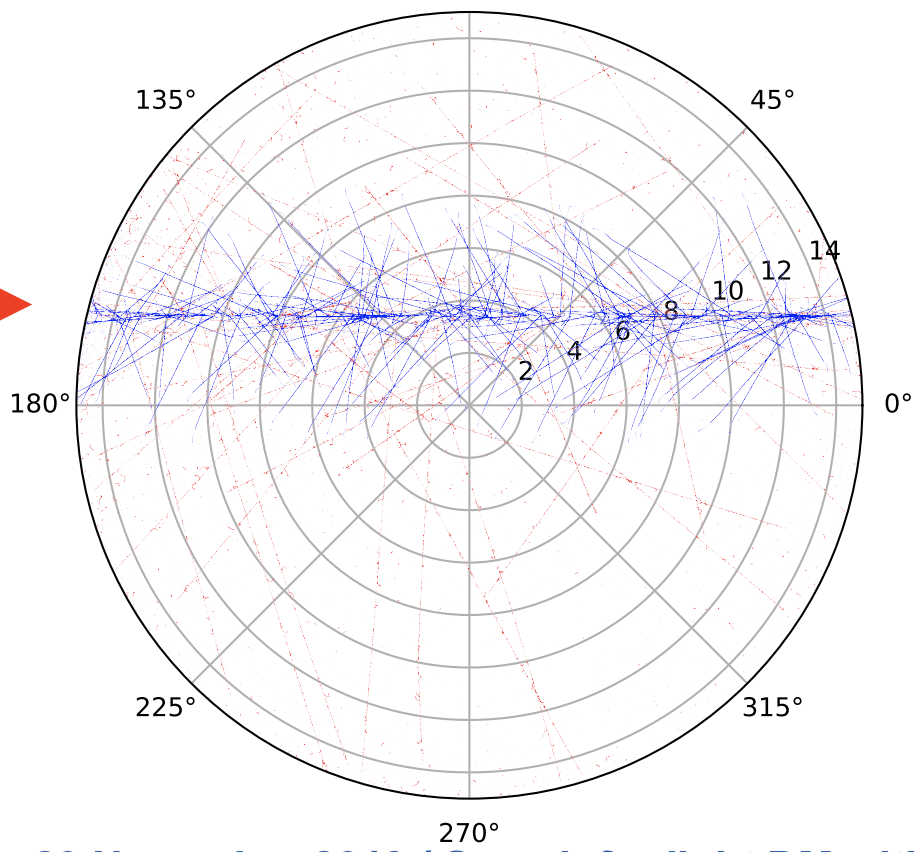}
\caption{Simulated events of 4\,MeV neutron interaction in the gaseous volume of the SPC and their reactions products. Protons tracks are depicted with red color while alpha particles with blue.\label{fig:simul}}

\end{figure}

The response of the detector is simulated using a framework developed at UoB based on GEANT4 and Garfield++ for high energy physics applications~\cite{Katsioulas:2019sui}. The simulation provides an expected efficiency for thermal neutrons of approximately $2.2\times 10^{-4}$, in good agreement with the measurements. Currently, the discrimination of $\alpha$-particle and proton events is being studied. This study is based on the different track lengths, as displayed in Fig.~\ref{fig:simul}, which can be translated into different pulse shape parameters. The model of the detector response will be used to unfold the measured energy spectra to recover the neutron energy spectra. Simulation results are currently being compared with measurements for validation.
\vspace*{-0.2cm}
\section{Neutron Measurements in Boulby and Future Applications}

The main goal is the deployment of the system in various applications including reactor monitoring, medical physics, accelerator facilities, homeland security, fuel searches, and space science. An initial application is the measurement of neutron background in underground laboratories, improving the sensitivity of experiments performing rare event searches.  A 30-cm in diameter aluminium SPC is installed in Boulby Underground Laboratory, currently performing fast neutron measurements with an $^{252}$Cf source. Also, measurements are planned in the UoB's MC40 cyclotron emulating hadron therapy environments. Detailed knowledge of neutron spectra are required to create a safer environment for patients and medical staff.
\vspace*{-0.2cm}
\section{Conclusions}

N$_{2}$SPHERE presents a reliable and affordable solution for the measurement of fast neutrons, using a large volume, nitrogen-filled SPC. The use of novel SPC instrumentation developments allows operation in high pressure, increasing sensitivity. The system has good efficiency for the detection of fast neutrons, without moderation, and limited influence from \textgamma-rays. Detailed simulations will be used to unfold neutron spectra from measurements. 

\addtolength{\textheight}{-12cm}   




\bibliographystyle{ieeetr}
\bibliography{references}
\end{document}